\documentclass[aps,prd,showpacs]{revtex4}
\usepackage{latexsym}
\usepackage{graphicx}% Include figure files
\begin{document}
% \draft

\thispagestyle{empty}

\title{Border of spacetime}
\author{Tomohiro Harada$^{1}$\footnote{Electronic
address: T.Harada@qmul.ac.uk},
Ken-ichi Nakao$^{2}$\footnote{Electronic
address: knakao@sci.osaka-cu.ac.jp}}
\affiliation{
$^{1}$Astronomy Unit, School of Mathematical Sciences, 
Queen Mary, University of London, 
Mile End Road, London E1 4NS, UK\\
$^{2}$Department of Physics, Osaka City University, 
Osaka 558-8585, Japan}
\date{\today}

\begin{abstract}                % DON'T CHANGE THIS LINE
It is still uncertain whether the cosmic censorship
conjecture is true or not.
To get a new insight into this issue, we propose 
the concept of 
the border of spacetime 
as a generalization of the spacetime 
singularity and discuss its visibility.
The visible border, corresponding to 
the naked singularity, is
not only relevant to 
mathematical completeness of general relativity
but also a window into new physics in strongly curved
spacetimes, which is in principle observable.
\end{abstract}

\pacs{04.20.Dw, 04.20.Cv}

\maketitle
%%---------------------------%

In general relativity, the singularity theorems 
tell us that 
spacetime singularities exist in generic gravitational 
collapse spacetime (see e.g., Ref.~\cite{he1972}). For singularities formed in 
gravitational collapse, Penrose~\cite{penrose1969,penrose1979} 
proposed the so-called 
cosmic censorship conjecture, which has two versions.
For spacetimes which contain physically reasonable matter fields
and develop from generic nonsingular initial data,
the weak one claims that there is no singularity
which is visible from infinity, while
the strong one claims that there is no singularity
which is visible to any observer.
A singularity censored by the strong version is
called a naked singularity, while a singularity
censored by the weak version is called 
a globally naked singularity.
There is no general proof for the conjecture at present.
Recent development on critical behavior 
(\cite{choptuik1993,mg2003}, see also Ref.~\cite{gundlach2003})
and self-similar attractor~(\cite{hm2001}, see also 
Ref.~\cite{harada2004}) 
has shown that 
there are naked-singular solutions
which result from nonsingular initial data 
and contain physically reasonable matter fields.
The critical solution has one unstable mode while 
the self-similar attractor has no unstable mode
against spherical perturbation,
although it is still uncertain whether 
these examples are stable against all other possible 
perturbations. 
If all examples of naked singularities were shown to be 
unstable, would it mean that they are all rubbish?

We have already known that general relativity 
will have
the limitation of its applicable scale in a high-energy side.
A simple and natural discussion on quantum effects of gravity yields 
the Planck energy $E_{\rm Pl}\sim10^{19}\mbox{GeV}$ 
as a cut-off scale $\Lambda$.
Some theories with large extra dimensions may have much lower 
cut-off scale, which could be
TeV scale~\cite{arkani-hamed,rs}. 
The energy scale of the curved spacetime can
be measured by the curvature
through Einstein's field equations. 
Then if the above expectation 
is true, general relativity 
is not applicable to the spacetime region whose 
curvature strength exceeds
$\Lambda^{4}/E_{\rm Pl}^{2}$.  
This consideration naturally leads to the notion of
the border of spacetime as follows.

Let $({\cal M},g)$ be a spacetime manifold $\cal M$ with a metric $g$. 
We call a spacetime region ${\cal A} \subset {\cal M}$
a {\it border}
if and only if the following inequality is satisfied:
\begin{equation}
\inf_{\cal A}F \ge {\Lambda^{4}\over E_{\rm Pl}^{2}}~,
\label{eq:definition}
\end{equation}
where the curvature strength $F$ is given, for instance, by
\begin{equation}
F:=\max(|R^{a}{}_{a}|,|R^{ab}R_{ab}|^{1/2},|R^{abcd}R_{abcd}|^{1/2}).
\end{equation}
We denote the union of all borders
in ${\cal M}$ by ${\cal U}_{\rm B}$. We call a border
$\cal A$ a {\it visible border} 
if and only if 
$J^{+}({\cal A},{\cal M})\cap ({\cal M}-{\cal U}_{\rm B})$ is not empty, 
where $J^{+}({\cal A},{\cal M})$ is the causal future of $\cal A$ in 
${\cal M}$~\cite{he1972}.

We can also naturally define a {\it globally visible border}.
To make the definition precise, we assume that   
the spacetime $\cal M$ is asymptotically flat and 
thus $({\cal M},g)$ is conformally embedded into a space 
$({\tilde {\cal M}},{\tilde g})$  as an unphysical 
spacetime manifold with boundary 
${\bar {\cal M}}={\cal M}\cup \partial {\cal M}$, where 
the boundary $\partial {\cal M}$ of $\cal M$ in $\tilde {\cal M}$ 
consists of the future and past null infinities 
${\cal I}^{+}$ and ${\cal I}^{-}$~\cite{he1972}. 
We call a border $\cal A$ a {\it globally visible border} if and only if 
$J^{+}({\cal A},{\bar {\cal M}})\cap {\cal I}^{+}$ is not empty. 
The asymptotic flatness implies that a globally visible border is 
a visible border since ${\cal I}^{+}$ is attached to a 
nonborder region of $\cal M$ in $\tilde {\cal M}$.
According to these definitions,
naked-singular spacetimes do not necessarily
involve visible borders. See Fig. 1 for an example
in which there is no visible border but naked singularity.

The presence of visible border 
implies the incompleteness of future predictability 
of general relativity. 
Even already known
examples of naked singularity formation 
will show that visible borders 
appear with nonzero probability in the collapse of physically reasonable 
matter fields. 
These visible borders will be stable against 
perturbations even if these perturbations prevent the formation of  
naked singularities. 
Even in the case of black hole formation, globally 
visible borders can appear. 
If the mass of the black hole $M_{\rm bh}$ is smaller than 
$E_{\rm Pl}^{3}/\Lambda^{2}$, the curvature strength in its exterior
can be larger than the cut-off 
scale $\Lambda^{4}/E_{\rm Pl}^{2}$ and hence it is necessarily 
a globally visible border. 
The critical behavior shows a possibility that such small 
black holes form in our universe within the framework of general 
relativity, and thus strongly suggests that
the cosmic censorship conjecture is violated 
in this physical sense. It is noted that
general relativity and other conventional physics are 
still applicable 
within the maximum Cauchy development in 
$({\cal M}-{\cal U}_{\rm VB})$, where ${\cal U}_{\rm VB}$ is 
the union of all visible borders.
Just on ${\cal U}_{B}$, the effects of new physics will
affect the spacetime metric directly or might invalidate
the notion of spacetime manifold itself, 
where the new physics may be quantum gravity or possibly 
the classical theories other than general relativity,
e.g. dilatonic gravity and higher dimensional theory.
In $({\cal M}-{\cal U}_{\rm B})$, the effects of new physics
will enter as boundary condtions.

However, if extremely severe finetuning of initial conditions 
is required for the formation of visible border, we might say 
that visible borders are
censored practically. We focus on this issue below.
First, if a naked-singular collapse solution is stable against 
all possible perturbations and an attractor,
no finetuning is needed for the formation 
of visible borders. 

Next, suppose a naked-singular collapse solution 
is a spherically symmetric self-similar solution with 
one unstable mode.
According to the renormalization group scenario~\cite{kha1995},
this solution will be identified with a critical solution.
Let $t$ and $r$ be appropriate time and radial coordinates, respectively,
and also $\tau\equiv -\ln(-Et)$ and $x\equiv \ln [Er/(-Et)]$, 
where $E(\ll \Lambda)$ is the characteristic energy scale at the 
initial moment. 
A general solution $h(\tau,x)$ is expressed by a trajectory 
in the space of initial data sets (SIDS), which is 
the space of functions of $x$, 
while a self-similar solution $h_{0}(x)$ corresponds to a 
fixed point.
If a fixed point has stable perturbations, 
there is a family of solutions asymptotically approaching 
the fixed point. 
This family will form a manifold embedded in the SIDS, 
which is called the stable manifold $S$ of the 
fixed point. 
For the case of the fixed point with one 
unstable mode, 
its stable manifold $S$ is codimension one 
in SIDS. Here let us consider a one-parameter family 
of initial data sets parametrized by $p$. This family 
has generically intersections with $S$ by virtue of the 
codimension of $S$ in SIDS. A value $p=p^*$ at an 
intersection corresponds to the critical value. The 
one-parameter family in the SIDS induces a one-parameter 
family of trajectories $h=h_p(\tau,x)$. Then only the 
trajectory $h=h_{p^*}(\tau,x)$ asymptotically approaches 
the fixed point $h=h_{0}(x)$ with one unstable mode, 
which corresponds to the critical solution.

For an initial data set with $p\approx p^{*}$,
the solution after a long time will be described by
\begin{equation}
h(\tau,x)\approx h_{0}(x)+(p-p^{*})e^{\kappa\tau}f_{\rm rel}(x),
\label{eq:deviation}
\end{equation}
where $\kappa$ ($>0$) and
$f_{\rm rel}$ are the eigenvalue and eigenfunction of the 
unstable mode, respectively. 
The first term $h_{0}(x)$ 
on the right-hand side in Eq.~(\ref{eq:deviation}) is 
dominant at first, and thus the solution initially shows almost 
self-similar behavior.
When the second term becomes comparable to 
the first one, the self-similar behavior is lost and 
the collapsing mass will start to form a black hole or bounce 
to disperse away. 
Let the mass within $x \leq x_{\rm bh}$ collapse to a 
black hole. Then the formation time $\tau=\tau_{\rm bh}$ of the black hole 
is estimated as $e^{\tau_{\rm bh}}
\sim |(p-p^*)f_{\rm rel}(x_{\rm bh}) 
/h_{0}(x_{\rm bh})|^{-\kappa^{-1}}=O(|p-p^{*}|^{-\kappa^{-1}})$, where 
$h_{0}(x_{\rm bh})$ and $f_{\rm rel}(x_{\rm bh})$ are of order unity.  
Using this result, we obtain the length scale $r=r_{\rm bh}$ 
of the collapsing mass at $\tau=\tau_{\rm bh}$, 
i.e.,  the gravitational radius 
$M_{\rm bh}/E_{\rm Pl}^{2}$ as 
\begin{equation}
{M_{\rm bh}\over E_{\rm Pl}^{2}}\sim r_{\rm bh}=E^{-1}e^{-\tau_{\rm bh}
+x_{\rm bh}}=O(E^{-1}|p-p^{*}|^{\kappa^{-1}}).
\end{equation}
Therefore
the black hole mass satisfies the above power-law behavior,
where the index $\gamma\equiv \kappa^{-1}$ is called the critical exponent.
The curvature strength $F$ is then estimated to be 
\begin{equation}
F\sim {E_{\rm Pl}^{4}\over M_{\rm bh}^{2}}=O(E^{2}
|p-p^*|^{-2\kappa^{-1}}).
\end{equation} 
From the above equation, the width of finetuning for the appearance 
of visible borders is estimated to be 
\begin{equation}
|p-p^{*}|=O\left(\left (\frac{\Lambda^2}
{E_{\rm Pl}E}\right)^{-\kappa}\right).
\end{equation}
The above estimate also applies to the subcritical case,
where the curvature strength reaches a maximum at the bounce 
and the matter field eventually disperses away.
If $\kappa$ is very small or equivalently $\gamma$ is very large, 
the width of finetuning is not too small even for 
$E\ll \Lambda^2/E_{\rm Pl}$.

For a naked-singular solution with $n$ unstable modes,
an $n$-parameter family of initial data sets has generically 
intersections with the stable manifold of codimension $n$. 
Then the finetuning 
should be considered in the $n$-dimensional parameter space.
Clearly, naked-singular solutions with fewer unstable
modes of small eigenvalues are physically more important.
Moreover, although in the above discussion we have supposed 
spherically symmetric perturbations, it is 
reasonably expected that
the discussion goes similarly even for
nonspherical perturbations~\cite{gundlach2002}.
We can also infer that the discussion also 
applies even to nonspherical naked-singular solutions.

Finally we discuss the detectability of visible borders in practice.
Let the mass $M$ be distributed nearly spherically and homogeneously 
within the length scale $L$. 
Then the curvature strength $F$ is 
typically estimated as $M/E_{\rm Pl}^{2}L^{3}$. 
If this mass is visible to an observer at 
infinity, $M/E_{\rm Pl}^{2}\lesssim L$ should be satisfied. 
This means that the curvature strength produced by the spherical 
visible mass satisfies $F\lesssim E_{\rm Pl}^4/M^{2}$. 
The definition (\ref{eq:definition}) of visible border 
implies $E_{\rm Pl}^{4}/M^{2}\gtrsim \Lambda^{4}/
E_{\rm Pl}^{2}$ 
so that the mass $M$ can form a spherically symmetric visible border. 
This means that the mass $M$ of the spherically symmetric visible border 
should satisfy
\begin{equation}
M \lesssim M_{\Lambda}:={E_{\rm Pl}^{3}\over \Lambda^{2}}.
\end{equation}
The upper bound $M_{\Lambda}$ is equal to or smaller than 
the lunar mass $\sim 10^{27}\mbox{g}\sim 10^{48}\mbox{erg}$ 
from the experimental constraint $\Lambda\agt 1 \mbox{TeV}$.
If the cut-off scale is much higher than TeV scale
and if the energy conservation law in a usual sense holds and
the mass of the visible border is not going to be negative,
the effect of one almost spherically symmetric visible border  
will be rather small as an energy source in astrophysical situations 
so that it may be difficult to astronomically 
observe the direct signal from inside the visible border.
In order that visible borders may be observable in 
a practical astronomical sense, 
the mass should be distributed in very specific manners,
e.g., a highly elongated visible border and
a loosely bound cluster of almost spherical 
visible borders, in an accordance with the hoop
conjecture~\cite{thorne1972}.
On the other hand, 
if recently proposed TeV scale gravity 
describes real gravitational physics at TeV scale,   
visible borders will be observed by the 
planned high-energy collider experiments
and/or as astrophysical high-energy phenomena.

The appearance of visible borders with nonzero probability 
implies not only the limitation of general relativity but 
also a new window into extremely high-curvature spacetime
physics in principle observable.

T.H. was supported from the JSPS.

\newpage
\begin{figure}[htbp]
\includegraphics[scale=0.5,angle=0]{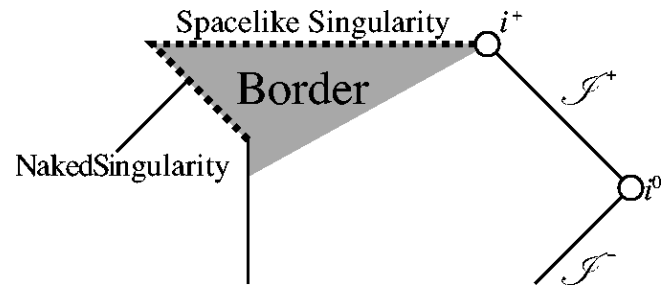}
\label{fig:Fig1}
\caption{Penrose diagram of a possible spacetime 
in which there is no visible border 
even if a naked singularity exists. The shaded region is a border region. 
The causal future of the border contains no nonborder region.}
\end{figure}

\end{document}